\documentclass[conference]{IEEEtran}
\usepackage{ifpdf}

\usepackage{cite}

\ifCLASSINFOpdf
   \usepackage[pdftex]{graphicx}
   \graphicspath{{../figures/}}
\else
  \usepackage[dvips]{graphicx}
  \graphicspath{{./figures/}}
\fi
\usepackage{amsmath}
\usepackage{url}

\usepackage{multirow}

\makeatletter
\def\ps@IEEEtitlepagestyle{
  \def\@oddfoot{\mycopyrightnotice}
  \def\@evenfoot{}
}
\def\mycopyrightnotice{
  {\footnotesize
  \begin{minipage}{\textwidth}
  \centering
  Copyright~\copyright~2017 IEEE.  Personal use of this material is permitted.  Permission from IEEE must be obtained for all other uses, in any current or future media, including reprinting/republishing this material for advertising or promotional purposes, creating new collective works, for resale or redistribution to servers or lists, or reuse of any copyrighted component of this work in other works.\\
DOI: 10.1109/INDIN.2017.8104836\\
URL: https://ieeexplore.ieee.org/document/8104836
  \end{minipage}
  }
}

\begin{document}
%
\title{ArChes - Automatic Generation of Component Fault Trees from Continuous Function Charts}

\author{\IEEEauthorblockN{Marc Zeller, Kai H{\"o}fig, Jean-Pascal Schwinn}
\IEEEauthorblockA{Siemens AG, Corporate Technology\\
Otto-Hahn-Ring 6, 81379 Munich, Germany\\
Email: \{marc.zeller, kai.hoefig, jean-pascal.schwinn\}@siemens.com}
}


%


\maketitle

\begin{abstract}
The growing size and complexity of software in embedded systems poses new challenges to the safety assessment of embedded control systems. In industrial practice, the control software is mostly treated as a black box during the system's safety analysis. The appropriate representation of the failure propagation of the software is a pressing need in order to increase the accuracy of safety analyses. However, it also increase the effort for creating and maintaining the safety analysis models (such as fault trees) significantly.
In this work, we present a method to automatically generate Component Fault Trees from Continuous Function Charts. This method aims at generating the failure propagation model of the detailed software specification. 
Hence, control software can be included into safety analyses without additional manual effort required to construct the safety analysis models of the software. Moreover, safety analyses created during early system specification phases can be verified by comparing it with the automatically generated one in the detailed specification phased.
\end{abstract}


%
\IEEEpeerreviewmaketitle

\section{Introduction}
\label{sec:introduction}
%
The importance of safety-critical software systems in many application domains of embedded systems, such as aerospace, railway, health care, automotive and industrial automation, is continuously growing. In order to guarantee the high quality demands in these application domains, also the effort for safety assessment is increasing. 
The goal of the safety assessment process is to identify all failures that cause hazardous situations and to demonstrate that their probabilities are sufficiently low. In the application domains of safety-critical systems the safety assurance process is defined by the means of safety standards (e.g.~the IEC 61508 standard \cite{iec61508}).
Traditionally, the analysis of a system in terms of safety consists of bottom-up safety analysis approaches, such as \textit{Failure Mode and Effect Analysis (FMEA)}, and top-down ones, such as \textit{Fault Tree Analysis (FTA)}, to identify failure modes, their causes, and effects with impact on the system safety.
With \emph{Component Fault Trees (CFTs)} \cite{Kaiser2003} there is a model- and component-based methodology for FTA, which supports a modular and compositional safety analysis strategy. Fault tree elements are related to their development artifacts and can be reused along with the respective development artifact.

In industry, software within safety-critical systems is currently increasing in size and importance. Hence, also the influence of software in safety analysis is increasing. However, in practice software is mostly treated as a black box within the safety analysis. 
The representation of the failure propagation of the software is a pressing need in order to increase the accuracy of the safety analyses, also the effort for creating and maintaining the safety analysis models is increasing significantly.
Moreover, in order to ensure the quality of the safety assessment manual and time-consuming reviews of the failure propagation model in terms of completeness and correctness are required.

In this work, we present a method to fully automatically generate Component Fault Trees from \textit{Continuous Function Charts (CFCs)}. This methodology aims at generating the failure propagation model of the detailed software specification automatically. 
Hence, safety analyses in form of Fault Tree Analysis (FTA), can be performed without manual effort required to construct the safety analysis models of the software. Moreover, the failure propagation model specified during system design can be verified by comparing it with the automatically generated one. 

There are a number of concepts to automatically generate failure propagation models from the system design. In \cite{bondavalli.1999,miguel.2008}
fault tree models are generated from UML models to perform safety analysis, while \cite{bretschneider.2004,papadopoulos01,rae.2004,szabo.2000} use a system architecture model, such as AADL, EAST-ADL, etc., as input to generate fault tree models.
Moreover, some approaches deal with the automated generation of failure propagation models from a data flow language such as the one used by Matlab/Simulink \cite{papadopoulos.2001.matlab,Tajarrod2008,BuonoRKZ15}.

However, all these approaches focus on the system architecture as input. Often the required manual modeling and preparation efforts are very high in order to be able to generate the failure propagation models.
In this work, we focus on the automatic generation of failure propagation models from a detailed software specification in form of continuous function charts.

The rest of the paper is organized as follows:
In Section \ref{sec:relatedwork} we briefly summarize relevant related work. Afterwards, we outline the concepts of CFCs in Section \ref{sec:sibas} and CFTs in Section \ref{sec:cfts}. Section \ref{sec:generation} presents our approach to automatically CFTs from CFCs.
The paper is concluded in Section \ref{sec:summary}.

%

\section{Continuous Function Charts}
\label{sec:sibas}
\textit{Continuous Function Chart (CFC)} is graphical programming language for \textit{Programmable Logic Controller (PLC)} to design complex control and regulation tasks as an extension of the IEC 61131-3 standard \cite{iec61131}. 
Instead of using a sequence of commands in textual notation, function blocks are combined and interconnected graphically. The CFC diagrams for programmable controller resemble electronic circuit diagrams. The function to be performed by the control system is represented in the form of the interconnected graphic elements. Since pre-defined function blocks only need to be connected to one another, complex functions can be programmed easily by developers coming from various engineering disciplines.

A CFC diagram consists of function blocks and linkages between these blocks. Each function block has different types of input and output parameters. It processes the input parameters according to a specific automation function and produces output parameters for other function blocks. The automation function of each function block is defined manually by a developer. The function blocks' outputs may be linked to the inputs of other function blocks in CFC diagrams. Thereby, each linkage indicates that an input parameter of a function block obtains its value from the specific output parameters of another function block.
%
%
Therefore, CFC diagrams have a precise syntax and each function block has a well defined semantics.
CFC diagrams can be created graphically using tools such as SIBAS.G or SIMATIC S7 among others.
SIBAS.G is an engineering tool for the development of software for the vehicle control of trains. SIMATIC S7 is used to design complex control-engineering tasks in the industrial automation domain.

\section{Component Fault Trees}
\label{sec:cfts}
A \emph{Component Fault Tree (CFT)} is a Boolean model associated to system development elements such as components \cite{Kaiser2003}. It has the same expressive power as classic fault trees \cite{vesely1981fault}. CFTs (as well as classic fault trees) are used to model the failure behavior of safety-critical systems. This failure behavior is used to document that a system is safe and can also be used to identify drawbacks of the design of a system.

In CFTs, a separate \emph{CFT element} is related to a component. Failures that are visible at the outport of a component are models using \emph{Output Failure Modes} which are related to a specific outport. To model how specific failures propagate from an inport of a component to the outport, \emph{Input Failure Modes} are used. The internal failure behavior that also influences the output failure modes is modeled using the Boolean gates such as \emph{OR} and \emph{AND} as well as \emph{Basic Events}. 


Every CFT can be transformed to a classic fault tree by removing the input and output failure modes elements.
The CFT model allows, additionally to the Boolean formula that are also modeled within the classic fault tree, to associate the specific failure modes to the corresponding ports where these failures can appear. 
By using CFT methodology, benefits during the development, such as an increased maintainability of the safety analysis model, can be observed \cite{hoefig2013}.

\section{Generation of CFTs from CFCs}
\label{sec:generation}
\begin{figure}
  \centering
  \includegraphics[width=8.2cm]{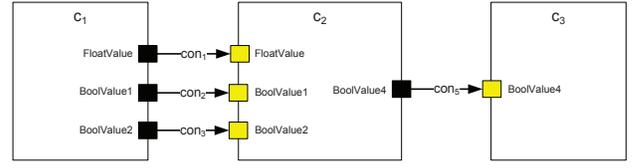}
  \caption{Exemplary system}
  \label{fig:example}
\end{figure}

In this section, we present our approach to automatically generate CFTs from continuous function charts (CFCs).
In the following, this method is described formally and illustrated using an example as depicted in Fig.~\ref{fig:example}.

Let the System $S$ consist of a set of components $C = \left\{ c_{1},...,c_{n} \right\}$. Each component $c \in C$ includes a set of inports $IN(c) = \left\{ in_{1},...,in_{p} \right\}$ and a set of outports $OUT(c) = \left\{ out_{1},...,out_{q} \right\}$.
The information flow between the outport of a component $c_i \in C$ and the inport of another component $c_j \in C$ (with $c_i \neq c_j$) of the system is represented as a set of connections
\begin{equation}
\forall c_i, c_j \in C: \; CON \subseteq OUT(c_i) \times IN(c_j)
\end{equation}

The example system as depicted in Fig.~\ref{fig:example} is defined by:
\begin{eqnarray*}
C &=& \left\{ c_1,c_2,c_3 \right\} \\
IN(c_1) &=& \emptyset \\
IN(c_2) &=& \{ FloatValue,BoolValue1,\\
				&& BoolValue2 \} \\
IN(c_3) &=& \left\{ BoolValue4 \right\} \\
OUT(c_1) &=& \{ FloatValue,BoolValue1,\\
				 && BoolValue2 \} \\
OUT(c_2) &=& \left\{ BoolValue4 \right\} \\
OUT(c_3) &=& \emptyset \\
CON &=& \{ (FloatValue,FloatValue),\\
		&& (BoolValue1,BoolValue1),\\
    && (BoolValue2,BoolValue2),\\
    && (BoolValue4,BoolValue4) \}
\end{eqnarray*}


The behavior of each component $c_i \in C$ is defined by a CFC diagram $cfc_i \in CFC$ with $\tilde{CFC}(c_i) = cfc_i$ and $cfc \neq \emptyset$.

Each CFC is defined by a tuple
\begin{equation}
cfc_i = \left( FB(cfc_i), LINK(cfc_i), IN(cfc_i), OUT(cfc_i) \right)
\end{equation}
where
$FB(cfc_i) = \left\{ fb_1,...,fb_m \right\}$ is a set of function blocks,
$LINK(cfc_i)$ is a set of linkages,
$IN(cfc_i) = IN(c_i)$ is a set of input parameters of the CFC and equals the set of inports of the corresponding component $c_i$, and
$OUT(cfc_i) = OUT(c_i)$ is a set of output parameters of the CFC and equals the set of outports of the corresponding component $c_i$.

A function block $fb_i \in FB(cfc_i)$ of a CFC $cfc_i \in CFC$ is defined as a tuple
\begin{equation}
fb_i = \left( t(fb_i), f(fb_i), IN(fb_i), OUT(fb_i) \right)
\end{equation}
where
$t(fb_i)$ is the unique type of a function block,
$f(fb_i)$ is the automation function,
$IN(fb_i) = \left\{ in_{i,1},...,in_{i,u} \right\}$ is a set of input parameters of the function block, and
$OUT(fb_i) = \left\{ out_{i,1},...,out_{i,v} \right\}$ is a set of output parameters of the function block.

A linkage $link_{j,i} \in LINK(cfc_i)$ of a CFC $cfc_i \in CFC$ is a relation
\begin{equation}
 \begin{split}
link_{j,i} = (( x_{k}, y_{l}) \; | \;& x_{k} \in OUT(fb_j) \cup IN(cfc_i), \\
																		& y_{l} \in IN(fb_i) \cup OUT(cfc_i) )
 \end{split}
\end{equation}
where $out_{j,k}$ is either the $k^{th}$ output parameter of function block $fb_j$ or the $k^{th}$ input parameter of the CFC and $in_{i,l}$ is either the $l^{th}$ input parameter of function block $fb_i$ or the $l^{th}$ output parameter of the CFC.

An automation function $f(fb_i)$ of a function block $fb_i \in FB(cfc_i)$ is a relation between its input and output parameters (e.g.~logical relations, such as \emph{or}, \emph{and}, \emph{not}, etc.). It is defined as $f(fb_i) \subseteq \left( IN(fb_i) \times OUT(fb_i) \right)$, where for all $in_{i,x} \in IN(fb_i)$ and $out_{i,y_1}, out_{i,y_2} \in OUT(fb_i)$, if $\left( in_{i,x}, out_{i,y_1} \right) \in f(fb_i)$ and $\left( in_{i,x}, out_{i,y_2} \right) \in f(fb_i)$ then $out_{i,y_1} = out_{i,y_2}$.

Each input parameter $in_{i,k} \in IN(fb_i)$ and output parameter $out_{i,l} \in OUT(fb_i)$ of a function block $fb_i \in FB(cfc_i)$ has a specific \emph{connector type} $CTY(x_{i}) = cty$ with $x_{i} \in IN(fb_i) \cup OUT(fb_i)$ (e.g.~Boolean, integer, etc.). If $link_{j,i} = (x_a, y_b) = (out_{fb_j,a}, in_{fb_i,b})$ then $CTY(out_{fb_j,a}) = CTY(in_{fb_i,b})$.

\begin{figure}
  \centering
  \includegraphics[width=8.5cm]{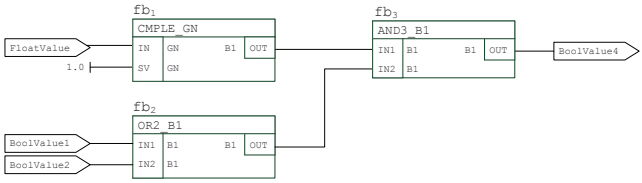}
  \caption{CFC diagram of component $c_2$ (in SIBAS.G)}
  \label{fig:cfc-example}
\end{figure}

In our example, the CFC diagram $cfc_2$ of component $c_2$ presented in Fig.~\ref{fig:cfc-example} is defined by:
\begin{eqnarray*}
FB(cfc_2)  &=& \left\{ fb_1,fb_2,fb_3 \right\} \\
IN(cfc_2)  &=& \{ FloatValue,BoolValue1,\\
						&& BoolValue2 \} \\
OUT(cfc_2) &=& \left\{ BoolValue4 \right\} \\
LINK(cfc_2)&=& \{ \left( FloatValue,IN_{1,1} \right), \\ 
            && \left( BoolValue1,IN1_{2,1} \right), \\
						&& \left( BoolValue2,IN2_{2,2} \right), \\
            && \left( OUT{1,1},IN1_{3,1} \right), \\
						&& \left( OUT_{2,1},IN2_{3,1} \right), \\
            && \left( OUT_{3,1},BoolValue4 \right) \} \\[3pt]
t(fb_1)    &=& CMPLE\_GN \\
IN(fb_1)   &=& \left\{ IN_{1,1}, SV_{1,2} \right\} \\
OUT(fb_1)  &=& \left\{ OUT_{1,1} \right\}
\end{eqnarray*}
\begin{eqnarray*}
t(fb_3)    &=& AND3\_Bl \\
IN(fb_3)   &=& \left\{ IN1_{3,1}, IN2_{3,2} \right\} \\
OUT(fb_3)  &=& \left\{ OUT_{3,1} \right\} \\[6pt]
CTY(IN_{1,1})  &=& CTY(SV_{1,2}) = GN \\
CTY(IN1_{2,1}) &=& CTY(IN2_{2,2}) = Bl \\
CTY(IN1_{3,1}) &=& CTY(IN2_{3,2}) = Bl \\
CTY(OUT_{1,1}) &=& CTY(OUT_{2,1}) \\
							 &=& CTY(OUT_{3,1}) = Bl
\end{eqnarray*}


If $c_i \in C$ has a CFT element $cft_i \in CFT$, then it is $\tilde{CFT}(c_i) = cft_i$ with $cft_i \neq \emptyset$.

Each CFT element $cft_i \in CFT(c_i)$ of a component $c_i \in C$ may have input failure modes
$IFM(in_k) = \left\{ ifm_1,...,ifm_s \right\}$ which are related each to an inport $in_k \in IN(c_i)$
as well as output failure modes $OFM(out_l) = \left\{ ofm_1,...,ofm_t \right\}$ which are related each to an outport $out_l \in OUT(c_i)$.

In order to specify the semantics of the failure modes within component fault tree an unambiguous failure type $fty$ is assigned to each input and output failure mode. The different failure types as well as the relation between them are specified in a so-called \textit{failure type system} $T$ \cite{318470}:
\begin{equation}
 \begin{split}
 FTY(fm) &= fty \\
  & \textnormal{ with } fm \in \bigcup_{i=1}^{p} IFM(in_i) \cup \bigcup_{j=1}^{q} OFM(out_j) \\
  & \textnormal{ and } fty \in T
 \end{split}
\end{equation}

Moreover, each CFT element $CFT(c_i) \neq \emptyset$ of a component $c_i \in C$ may have a set of gates $G = \left\{ g_1,...,g_r \right\}$.
Each gate $g_i \in G$ has exactly one output $g_i.out$, one or more inputs $g_i.IN = \left\{ g_i.in_1,...,g_i.in_s \right\}$, and a Boolean formula $b$ (e.g.~$g.out = g.in_1 \vee g.in_2$).

Input and output failure modes as well as gates are connected by a set of directed edges
\begin{equation}
 \begin{split}
E \subseteq \{ (out_x, in_y) \; | \;& out_x \in \bigcup_{i=1...p} IFM(in_p) \bigcup_{j=1...r} g_j.out \\
																		& in_y \in \bigcup_{k=1...r} g_k.IN \bigcup_{l=1...q} OFM(out_l) \}
 \end{split}
\end{equation}


The generation of a CFT from a CFC diagram is performed in three steps, which are defined as follows:

\subsection{Generation of CFT elements}
At first, a CFT element is created for each CFC diagram within a specific project:
\begin{equation}
\forall \; cfc_i \in CFC \text{ with } cfc_i = \tilde{CFC}(c_i): \; \tilde{CFT}(c_i) = cft_i 
\end{equation}
Thus, $\forall c_i \in C: \; \exists \; cft_i \in CFT$.

Moreover, based on the inputs and outputs defined in each CFC diagram, inports and outports are generated and interconnected based on the unique names of the inputs and outputs of the CFC diagrams:
\begin{equation}
\forall \; cfc_i \in CFC: \; IN(c_i) \rightarrow IN(cfc_i) 
\end{equation}
\begin{equation}
\forall \; cfc_i \in CFC: \; OUT(c_i) \rightarrow OUT(cfc_i)
\end{equation}
and
\begin{equation}
\begin{split}
\forall \;& cfc_i, cfc_j \in CFC \textrm{ with } cfc_i \neq cfc_j: \\
					& \; \forall \; out_{i,k} \in OUT(cfc_i), in_{j,l} \in IN(cfc_j): \\
					& \; \rightarrow \left\{ \left(out_{i,k},in_{j,l}\right) \; | \; name(out_{i,k}) = name(in_{j,l}) \right\}
\end{split}
\end{equation}

For the exemplary system as depicted in Fig.~\ref{fig:example} \& \ref{fig:cfc-example}, the following CFT elements are generated (see Fig.~\ref{fig:generatedCFTelement1}:
\begin{eqnarray*}
\tilde{CFT}(c_1) &=& cft_1 \\
\tilde{CFT}(c_2) &=& cft_2 \\
\tilde{CFT}(c_3) &=& cft_3 \\[3pt]
IN(cfc_2)  &=& \{ FloatValue,BoolValue1,\\
					 && BoolValue2 \} \\
OUT(cfc_2) &=& \left\{ BoolValue4 \right\} \\
CON &=& \{ (FloatValue,FloatValue), \\
		&& (BoolValue1,BoolValue1), \\
    && (BoolValue2,BoolValue2), \\
    && (BoolValue4,BoolValue4) \}
\end{eqnarray*}

\subsection{Generation of Input \& Output Failure Modes}
In the next step, the input and output failure modes are generated for each of the previously created CFT elements.

The generation of the failure modes is based on a generic mapping between the connector types in the CFC diagram and the failure types of the failure modes in the CFT element. Whereas, each connector type corresponds to a set of failure types from the generic failure type system $T$ \cite{318470}:
\begin{equation}
MAP: \; CTY(x_i) \mapsto \left\{ fty_1,...,fty_n \right\} \in T
\end{equation}
with $x_i \in IN(cft_i) \cup OUT(cft_i)$ and $cft_i \in CFT$ and $fty_j \in T$.

The generic mapping from connector types in CFCs to failure types in CFTs is presented in Table \ref{tab:DatatypesInSIBASG}.
\begin{table}[!h]
	\begin{center}
    \begin{tabular}{|c|l|}
    \hline
    \textbf{Connector Type} & \textbf{Failure Type} \\
    \hline
    \multirow{6}{*}{Boolean}              & false positive,\\
                                          & false negative,\\
                                          & omission,\\
                                          & commission,\\
                                          & too early,\\
                                          & too late\\
    \hline
    \multirow{6}{0.9cm}{Integer, Float, Time} & too high, \\
                                          & too low, \\
                                          & omission,\\
                                          & commission,\\
                                          & too early,\\
                                          & too late\\
    \hline
    \end{tabular}
	\end{center}
  \caption{Mapping of connector types in CFCs to failure modes in CFT}
	\label{tab:DatatypesInSIBASG}
\end{table}
For each CFT element $cft_i \in CFT$, a set of input failure modes as well as a set of output failure modes is generated based on the connector types of the inputs $IN(cfc_i)$ and outputs $OUT(cfc_i)$ of the corresponding CFC diagram $cfc_i \in CFc$, where $cfc_i = \tilde{CFC}(c_i)$ and $cft_i = \tilde{CFT}(c_i)$ with $c_i \in C$:
\begin{equation}
\begin{split}
\forall \;& cfc_i \in CFC: \; \forall \; in_j \in IN(cfc_i): \\
     & \rightarrow \left\{ ifm_k \; | \; MAP(CTY(in_j)) = FTY(ifm_k) \right\}
\end{split}
\end{equation}
and
\begin{equation}
\begin{split}
\forall \;& cfc_i \in CFC: \; \forall \; out_j \in OUT(cfc_i): \\
     & \rightarrow \left\{ ofm_k \; | \; MAP(CTY(out_j)) = FTY(out_k) \right\}
\end{split}
\end{equation}

\begin{figure}[!htp]
  \centering
  \includegraphics[width=8.5cm]{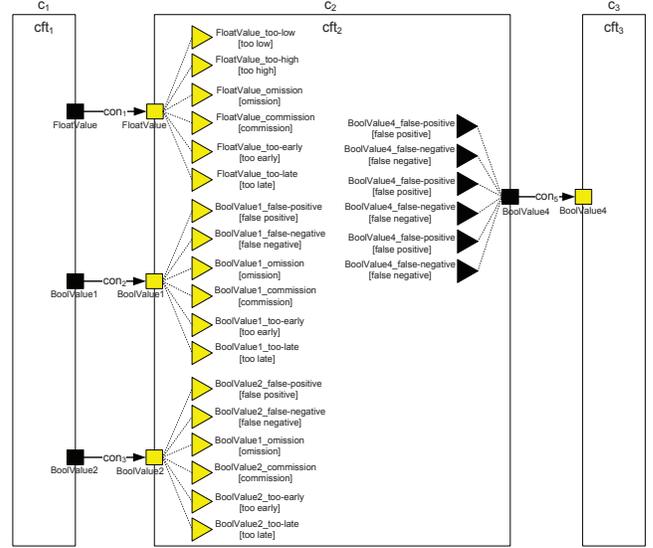}
  \caption{Exemplary system: Generated input and output failure modes}
  \label{fig:generatedCFTelement1}
\end{figure}

For the component $c_2$ of the exemplary system presented in Fig.~\ref{fig:example} and Fig.~\ref{fig:cfc-example}, a set of input and output failure modes are generated as depicted in Fig.~\ref{fig:generatedCFTelement1} where each input or output failure mode has a specific failure type $FTY$ (displayed in square brackets).

\subsection{Generation of the failure propagation}
In a last step, the failure propagation from the input failure modes of each CFT element $cft_i \in CFT$ to its output failure modes is generated based on the definition of the corresponding CFC diagram $cfc_i = \tilde{CFC}(c_i) \in CFC$. Therefore, input and output failure modes of the CFT element $cft_i$ are connected using Boolean gates \cite{Kaiser2003}.

At first, a set of Boolean gates $G$ is generated based on specific predefined rules for each function block $fb_j \in FB(cfc_i)$. Therefore, a set of \emph{rules} $R(t(fb_i)) = \left\{ r_1,...,r_s \right\}$ must be defined for each type of function block $t(fb_j) \in FB(cfc_i)$ of the CFC. It describes for all possible failure types how the output parameter of a function block are related to the possible failure types its input parameters (see Eq.~\ref{eq:map}).
\begin{figure*}[!t]
\normalsize
\begin{equation}
  \begin{split}
  \forall \;& cfc_i \in CFC: \; \forall \; fb_j \in FB(cfc_i): \; \forall \; out_{k,l} \in OUT(fb_j): \\
  & R(t(fb_j)) = \left\{ r_i \; | \; r_i:  MAP(CTY(out_{k,l})) \mapsto MAP(CTY(in_{k,1})) \circ \ldots \circ MAP(CTX(n_{k,u}) \right\} \\
  & \textrm{ with } \circ = \left\{ \neg,\wedge,\vee,\oplus \right\} \\
  & \textrm{ and } i = 1,..., \left|MAP(CTY(out_{k,l}))\right|
  \end{split}
	\label{eq:map}
\end{equation}
\end{figure*}
The possible failure types of the input and output parameters of a function block in the CFC are defined by their connector type according to the mapping $MAP$ (see Table \ref{tab:DatatypesInSIBASG}).
For instance, if the connector type of the output parameter is a \emph{Boolean}, two rules must be defined: one for the failure type \emph{false negative} and one rule for the failure type \emph{false positive}.

In case that no rule is predefined for a type of function block $fb_j \in FB(cfc_i)$ used in the CFC diagram $cfc_i \in CFC$, only the worst case scenario for the failure propagation can be assumed. This worst case scenario is defined in Eq.~\ref{eq:worstcase}.
\begin{figure*}[!t]
\normalsize
\begin{equation}
  \begin{split}
  \forall \;& out_{k,l} \in OUT(fb_j): \\
  & R(\textrm{worst case}) = \left\{ r_i \; | \; r_i: MAP(CTY(out_{k,l})) \bigvee_{s=1}^{u} \bigvee MAP(CTY(in_{k,t})) \right\} \\
  & \textrm{ with } i = 1,..., \left|MAP(CTY(out_{k,l}))\right|
  \end{split}
	\label{eq:worstcase}
\end{equation}
\end{figure*}

For the function blocks of component $c_2$ in our example as depicted in Fig.~\ref{fig:cfc-example} the following rules are defined:
\begin{eqnarray*}
\begin{aligned}
R(&CMPLE\_GN) = \\
							& \{ OUT.false\mbox{-}positive = IN.too\mbox{-}low, \\
              &\; OUT.false\mbox{-}negative = IN.too\mbox{-}high, \\
              &\;  OUT.omission = IN.omission, \\
              &\;  OUT.commission = IN.commission, \\
              &\;  OUT.too\mbox{-}early = IN.too\mbox{-}early, \\
              &\;  OUT.too\mbox{-}late = IN.too\mbox{-}late \} \\[3pt]
R(&OR2\_Bl) = \\
							& \{ OUT.false\mbox{-}positive = IN1.false\mbox{-}positive \; \vee \\[-3pt]
							&\;\qquad\qquad\qquad\qquad\qquad\; IN2.false\mbox{-}positive, \\
              &\;  OUT.false\mbox{-}negative =  IN1.false\mbox{-}negative \; \wedge \\[-3pt]
							&\;\qquad\qquad\qquad\qquad\qquad\; IN2.false\mbox{-}negative, \\
              &\;  OUT.omission = IN1.omission \; \wedge \\[-3pt]
							&\;\qquad\qquad\qquad\qquad IN2.omission,\\
              &\;  OUT.commission = IN1.commission \; \vee \\[-3pt]
							&\;\qquad\qquad\qquad\qquad\quad\; IN2.commission,\\
              &\;  OUT.too\mbox{-}early = IN1.too\mbox{-}early \; \vee \\[-3pt]
							&\;\qquad\qquad\qquad\quad\;\;\; IN2.too\mbox{-}early,\\
              &\;  OUT.too\mbox{-}late = IN1.too\mbox{-}late \; \wedge \\[-3pt]
							&\;\qquad\qquad\qquad\quad\; IN2.false\mbox{-}late \} \\[3pt]
R(&AND3\_Bl) = \\
							& \{ OUT.false\mbox{-}positive = IN1.false\mbox{-}positive \; \wedge \\[-3pt]
							&\;\qquad\qquad\qquad\qquad\qquad\; IN2.false\mbox{-}positive, \\
              &\;  OUT.false\mbox{-}negative = IN1.false\mbox{-}positive \; \vee \\[-3pt]
							&\;\qquad\qquad\qquad\qquad\qquad\;IN2.false\mbox{-}positive, \\
              &\;  OUT.omission = IN1.omission \wedge \\[-3pt]
							&\;\qquad\qquad\qquad\qquad IN2.omission, \\
              &\;  OUT.commission = IN1.commission \; \vee \\[-3pt]
							&\;\qquad\qquad\qquad\qquad\quad\; IN2.commission, \\
              &\;  OUT.too\mbox{-}early = IN1.too\mbox{-}early \; \vee \\[-3pt]
							&\;\qquad\qquad\qquad\quad\quad IN2.too\mbox{-}early, \\
              &\;  OUT.too\mbox{-}late = IN1.too\mbox{-}late \; \wedge \\[-3pt]
							&\;\qquad\qquad\qquad\quad\; IN2.false\mbox{-}late \}
\end{aligned}
\end{eqnarray*}
Based on these rules, the following Boolean gates are generated for the CFT element $cft_2$ of component $c_2$:
\begin{equation*}
\begin{split}
AND1\mbox{-}1 &= AND1\mbox{-}2 = AND1\mbox{-}3 \\
							&= AND2\mbox{-}1 = AND2\mbox{-}2 = AND2\mbox{-}3 \\
							&= ( AND.out, \left\{ AND.in_1, AND.in_2 \right\}, \\
              & \qquad AND.out = AND.in_1 \wedge AND.in_2 ) \\
OR1\mbox{-}1 &= OR1\mbox{-}2 = OR1\mbox{-}3 \\
						 &= OR2\mbox{-}1 = OR2\mbox{-}2 = OR2\mbox{-}3 \\
						 &= ( OR.out, \left\{ OR.in_1, OR.in_2 \right\}, \\
             & \qquad OR.out = OR.in_1 \vee OR.in_2 )
\end{split}
\end{equation*}

Afterwards, the input and output failure modes as well as the Boolean gates of each CFT element $cft_i \in CFT$ are interconnected according to the corresponding CFC's linkage $LINK(cfc_i)$. For creating the interconnections, we start from the output failure modes and connect them with the available input failure modes through the Boolean gates. 
Therefore, a set of direct edges is created as follows:
\begin{equation}
  \begin{split}
    \forall \;& out_j \in OUT(cfc_i): \; \forall \; \forall ofm_k \in OFM(out_j): \\
    &\rightarrow \{ \left(x,ofm_k\right) \; | \; x \in IFM(in_l) \vee x = g_r.out, \; \\
    &\qquad\quad\; \exists \;link_z \in LINK(cfc_i) = \left( y, OUT(cfc_i) \right), \\
		&\qquad\quad\quad y = in_l \vee y \cup OUT(fb_d), fb_s \Rightarrow g_r
    \}
  \end{split}
\end{equation}
and
\begin{equation}
  \begin{split}
    \forall \;& g_j.IN \in G: \; \forall \; g_j.in_k \in g_j.IN: \\
    &\rightarrow \{ \left(x,g_j.in_k\right) \; | \; x \in IFM(in_l) \vee x = g_r.out, g_r \neq g_j \; \\
    &\qquad\qquad\; \exists \;link_z \in LINK(cfc_i) = \left( y, OUT(cfc_i) \right), \\
		&\qquad\qquad\quad y = in_l \vee y \cup OUT(fb_s), fb_s \Rightarrow g_r
    \}
  \end{split}
\end{equation}

\begin{figure*}[htp]
  \centering
  \includegraphics[width=9.5cm]{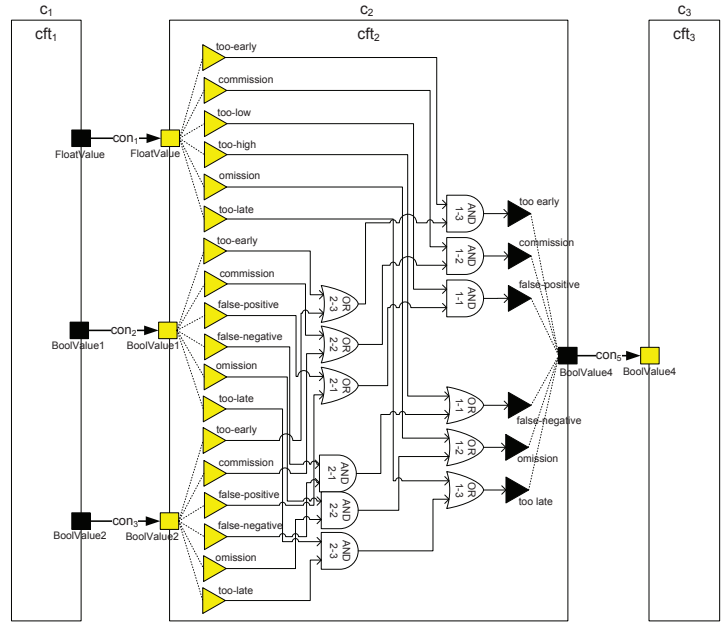}
  \caption{Exemplary system: Generated CFT element for component $c_2$}
  \label{fig:generatedCFTelement2}
\end{figure*}

The result of the generation process for the exemplary system is presented in Fig.~\ref{fig:generatedCFTelement2}.

\section{Conclusions}
\label{sec:summary}
%
Our approach to generate Component Fault Trees (CFTs) from Continuous Function Charts (CFCs) automatically creates a failure propagation model based on thte detailed software specification of the system. No additional manual effort is needed to perform a safety analysis.
The resulting CFT can be used for Fault Tree Analyses (FTA) of the overall system including the software as a white box. Thus, the accuracy of the safety analysis is increased without additional effort needed for the construction and maintenance of the safety analysis model for the controller software of the system.
Compared to traditionally used methods for the analysis of software such as root cause analysis or fault injection tests, in which the quality of the results is depending on the accuracy of the input data which are generated manually by a team of experts (e.g.~by brainstorming), our approach provides a complete set of input data while no manual effort is required. 
Moreover, the automatically generated CFT may be compared to a CFT (or Fault Tree) created manually by a safety engineering during the system specification (e.g.~as described by \cite{5635142}). Hence, it is possible to verify if the failure propagation model specified during system design is build correctly in terms of consistency and completeness.

%


\bibliography{paper_references}
\bibliographystyle{IEEEtran}

\end{document}